\documentclass[journal]{IEEEtran}

\usepackage{algorithm}
\usepackage{algpseudocode}
\usepackage{pifont}

\usepackage{moreverb}

\usepackage[colorlinks,bookmarksopen,bookmarksnumbered,citecolor=red,urlcolor=red]{hyperref}

\usepackage[tight,footnotesize]{subfigure}
\usepackage{latexsym}
\usepackage{amssymb}
\usepackage{amsfonts}
\usepackage{amsthm}
\usepackage{amsmath}
\usepackage{graphicx}
\usepackage{array}
\usepackage{inputenc}
\usepackage{tikz}
\usepackage{textcomp}
\usepackage{color,soul}
\usepackage{tabulary}
\usepackage{booktabs}

\usepackage{flushend}

\usepackage{pifont}

\begin{document}

\title{Multi-cell Device-to-Device Communications: \\A Spectrum Sharing and Densification Study}

\author{\IEEEauthorblockN{Marios~I.~Poulakis\IEEEauthorrefmark{1},
 Antonis~G.~Gotsis\IEEEauthorrefmark{1}\IEEEauthorrefmark{2},
 and Angeliki~Alexiou\IEEEauthorrefmark{1}}\\
\IEEEauthorblockA{\IEEEauthorrefmark{1}Department of Digital Systems, University of Piraeus, Greece} \\
\IEEEauthorblockA{\IEEEauthorrefmark{2}Feron Technologies P.C., Greece\\
Emails: \{mpoulak, agotsis, alexiou\}@unipi.gr}}


\maketitle

\begin{abstract}
One of the most significant 5G technology enablers will be Device-to-Device (D2D) communications. D2D communications constitute a promising way to improve spectral, energy and latency performance, exploiting the physical proximity of communicating devices and increasing resource utilization. Furthermore, network infrastructure densification has been considered as one of the most substantial methods to increase system performance, taking advantage of base station proximity and spatial reuse of system resources. However, could we improve system performance by leveraging both of these two 5G enabling technologies together in a multi-cell environment? How does spectrum sharing affect performance enhancement? This article investigates the implications of interference, densification and spectrum sharing in D2D performance gain. The in-band D2D approach, where legacy users coexist with potential D2D pairs, is considered in a multi-cell system. Overlay and underlay spectrum sharing approaches are employed in order for the potential D2D pairs to access the spectrum. Given that two of the most critical problems in the D2D concept are mode selection and user scheduling, we jointly address them, aiming at maximizing the total system uplink throughput. Thus, we present a radio resource management mechanism for intra-cell and cross-cell overlay/underlay D2D communications enabled in a multi-cell system. System-level simulations are executed to evaluate the system performance and examine the trends of D2D communication gain for the different spectrum sharing approaches and various densification scenarios. Finally, real-world SDR-based experiments are performed to test and assess D2D communications for overlay and underlay spectrum sharing.
\end{abstract}


\IEEEpeerreviewmaketitle

\section{Introduction}
5G wireless systems are expected to boost network capacity, spectral and energy efficiency, peak data rates, number of connected devices and consequently mobile data volumes with seamless and ubiquitous ultra-low latency connections. One of the key technology components of the evolving 5G architecture \cite{Lien}-\cite{METIS} will be the Device-to-Device (D2D) communications, which refer to the capability of direct communication between two or more devices without the intervention of a base station. Recently, D2D communications have attracted strong attention in academia and industry \cite{Mach}-\cite{Hwang}, initially intended for public safety scenarios \cite{Lien, 3GPP36.843, 3GPP36.877}, however other user-oriented (social applications) and network-oriented (offloading) use-cases have been rapidly emerged.\newline
\indent D2D communication is a promising way to improve performance providing different types of gain: proximity gain, hop gain and reuse gain \cite{Fodor}. However, important challenges are raised. Particularly, efficient resource sharing and radio protocol design should be proposed \cite{Mach}, in order to address critical D2D processes such as mode selection, scheduling and discovery, avoiding properly the interference to legacy cellular users (CUEs). D2D communication can be classified into in-band and out-band. The in-band D2D communication model refers to the case where D2D communications take place in a licensed spectrum allocated to the cellular operators. Uplink (UL), downlink (DL), or both resources can be reused. On the other hand, the out-band D2D utilizes the unlicensed spectrum adopted by other wireless technologies (e.g. Wi-Fi, Bluetooth). \newline
\indent In this article, the in-band D2D communication model is employed, where D2D communications take place in the UL spectrum allocated to the cellular operators, with the D2D users (DUEs) to be able to access this spectrum in a dedicated (overlay or orthogonal) or a shared (undelay or non-orthogonal) way. Overlay D2D communication avoids the interference to legacy CUEs issue, because D2D and cellular resources do not overlap, while the D2D management is handled by the cellular operators. On the contrary, in underlay D2D communications, DUEs and legacy CUEs share the same radio resources generating interference among each other, while D2D can be fully, loosely or not controlled by cellular operators. In both approaches, new resource allocation and synchronization methods should be introduced and significant changes in the existing standard are needed. It is worthwhile to mention that the 3GPP is directed to in-coverage or partial-coverage scenarios \cite{3GPP36.213, 3GPP36.331}, where D2D communication utilizes UL resources for the sidelink \cite{Lien}. The argument for the UL use is that this direction is mostly underutilized compared to the DL and the interference situation is easier to be resolved because the “victim” of D2D interference is evolved NodeB (eNB).\newline
\indent Furthermore, network infrastructure densification has been proposed as one of the leading concepts to cope with the growing traffic trends \cite{Gotsis, Kamel}. The basic idea is to get base stations and access points (having small transmission power) as close as possible to the end users. Consequently, the spectrum is increasingly reused, improving system capacity, and the link to the end user becomes shorter improving link quality. However, the network densification cannot continue improving performance endlessly and the question of what are its fundamental limits was addressed in \cite{Andrews}.\newline
\indent Given that one of the components of D2D communication benefit is proximity gain, a reasonable question raised is how network densification affects the D2D performance in D2D-enabled systems. Since a more dense network means that end users are closer to base station, a densification threshold may exist and above that D2D communication will not remain beneficial. Toward this end, this article carefully examines the effects of densification and spectrum sharing on the D2D performance. More specifically, a radio resource management (RRM) mechanism is introduced to jointly handle the D2D mode selection and user (UE) scheduling in a multi-cell environment considering intra-cell and cross-cell D2D capabilities. These two important procedures should be addressed jointly, since they are highly intertwined. In such an environment, the interference and densification are expected to significantly affect the D2D gain, while it is important to take into account the cross-cell D2D communications, especially  for cell-edge users cases. The proposed RRM mechanism is separately designed for the overlay and the underlay spectrum sharing approaches. The joint optimal policy is quite complex, thus a problem reformulation is needed. \newline
\indent A system-level simulator integrating legacy CUEs along with D2D capable UEs is developed to evaluate the performance of D2D communications and examine the trends of total D2D communication gain (defined as the combination of direct gain and offloading gain) for various system parameters. Finally, the performance of in-band D2D communications for overlay and underlay spectrum sharing is experimentally evaluated in a Software Defined Radio (SDR)-based joint LTE-D2D implementation.

\section{Multi-cell D2D Communications Design: Mode Selection, Spectrum Sharing \& Densification}
\subsection{Design Overview}
\begin{figure}[!t]
\centering{
\includegraphics[width=3.4in, scale=1]{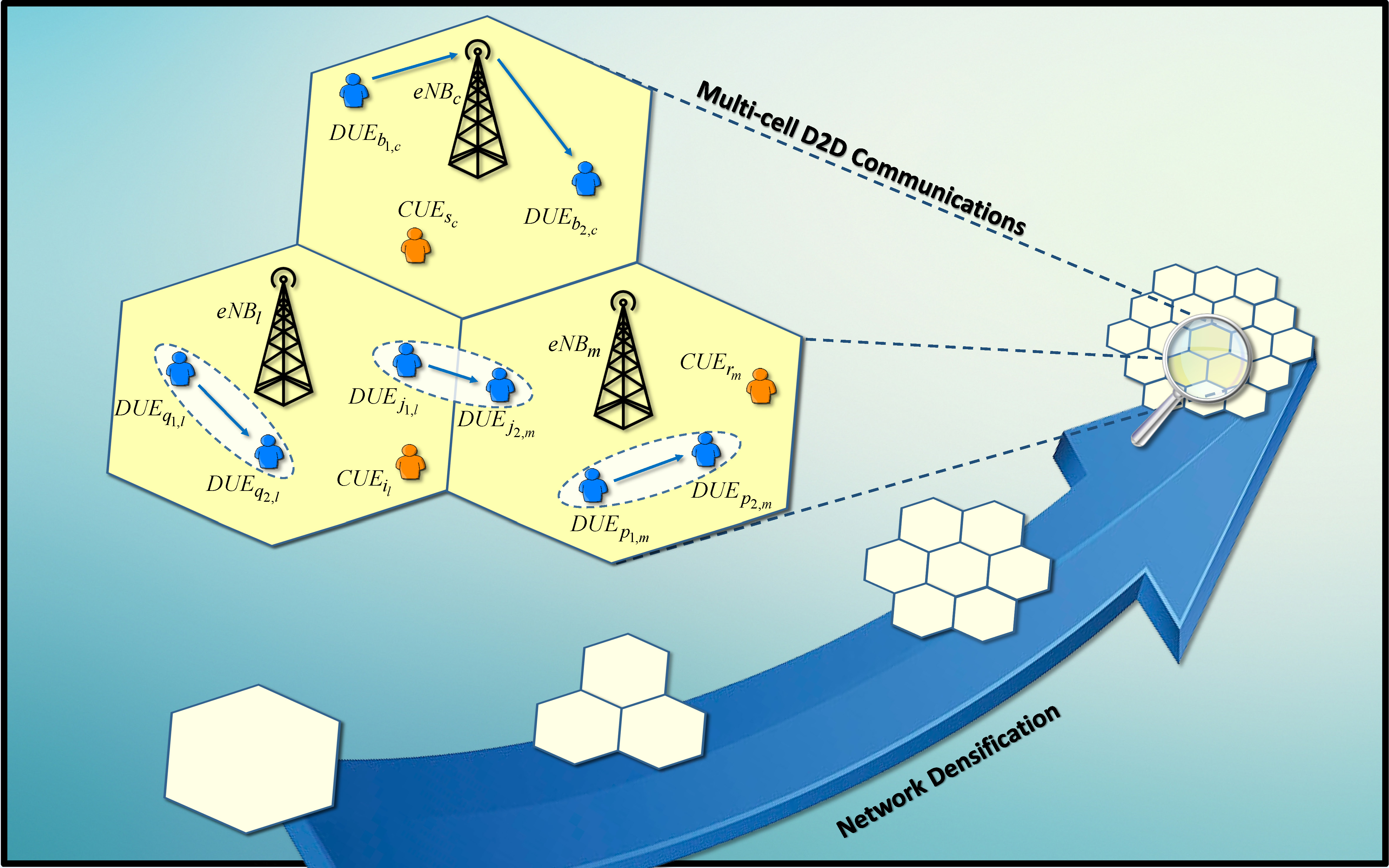}}
\caption{Multi-cell D2D communications \& densification.}
\label{fig1}
\end{figure}

Several D2D communication scenarios have been specified \cite{Mach}, depending on the coverage of cellular network (i.e. in coverage, partial coverage, out of coverage), the type of D2D communication (i.e. one-to-one, one-to-many), the area of D2D communication (i.e. same cell, different cell) and the relaying functionality (i.e. to enhance capacity, to extend coverage). However, regardless of the application scenario, the design of multi-cell D2D-enabled systems faces many technical challenges, such as mode selection, scheduling, interference management, synchronization and power control. In this article, we focus on the mode selection and scheduling procedures, while we examine different spectrum sharing approaches according to which cellular resources are reused for D2D communication (resulting in different interference scenarios between D2D and cellular users). Furthermore, we study the role of network densification in the design of D2D-enabled cellular system. \newline
\indent Fig.~\ref{fig1} illustrates a multi-cell wireless network with D2D communication capabilities, consisting of $K^{leg}$ legacy CUEs, where each UE $i_l \in {\cal K}_l^{leg}= \{ 1,2,...,{K^{leg}}\}$ is located in cell $l \in {\cal L}= \{ 1,2,...,L\}$, and $K^{pD2D}$ potential D2D pairs, where each pair ${j_{l - m}} = [j_{1,l},j_{2,m}] \in {{\cal K}^{pD2D}}= \{ 1,2,...,{K^{pD2D}}\}$ has the transmitter $DUE_{j_{1,l}}$ in cell $l \in {\cal L}$ and the receiver $DUE_{j_{2,m}}$ in $m \in {\cal L}$, in the general case (intra- or cross-cell D2D). Note that ${\cal K} = {{\cal K}^{leg}} \cup {{\cal K}^{pD2D}} = \{ 1,2,...,K\}$, where $K = K^{leg} + 2K^{pD2D}$ is the total number of UEs in the system. Potential D2D pairs can access the spectrum using in-band overlay or underlay spectrum sharing and the considered scenario represents the in-coverage scenario with same or different cell D2D capabilities. The UEs are stationary and uniformly distributed in the hexagonal cells of $L$ eNBs. \newline
\subsection{Mode Selection}
\indent The communication mode of the potential D2D pairs specifies if the DUEs communicate directly with each other or via the eNB. Moreover, together with the selected spectrum sharing scheme, it identifies whether DUEs utilize the same radio resources as the conventional cellular communication or not. Thus, proper mode selection plays an important role in D2D communication, which can be achieved through the following available transmission modes:
\begin{itemize}
  \item {\it{Cellular Mode (CM)}}, where DUEs of a potential D2D pair communicate through their associated eNB(s), i.e. $DUE_{j_{1,l}} \rightarrow eNB_{l} (\rightarrow eNB_{m}) \rightarrow DUE_{j_{2,m}}$,
  \item {\it{Direct or D2D Mode (DM)}}, where a potential D2D pair becomes an actual D2D pair whose DUEs communicate directly, i.e.  $DUE_{j_{1,l}} \rightarrow DUE_{j_{2,m}}$.
\end{itemize}
\subsection{Spectrum Sharing}
\begin{figure}[!t]
\centering{
\hspace*{-3mm}
\includegraphics[width=3.7in, scale=1]{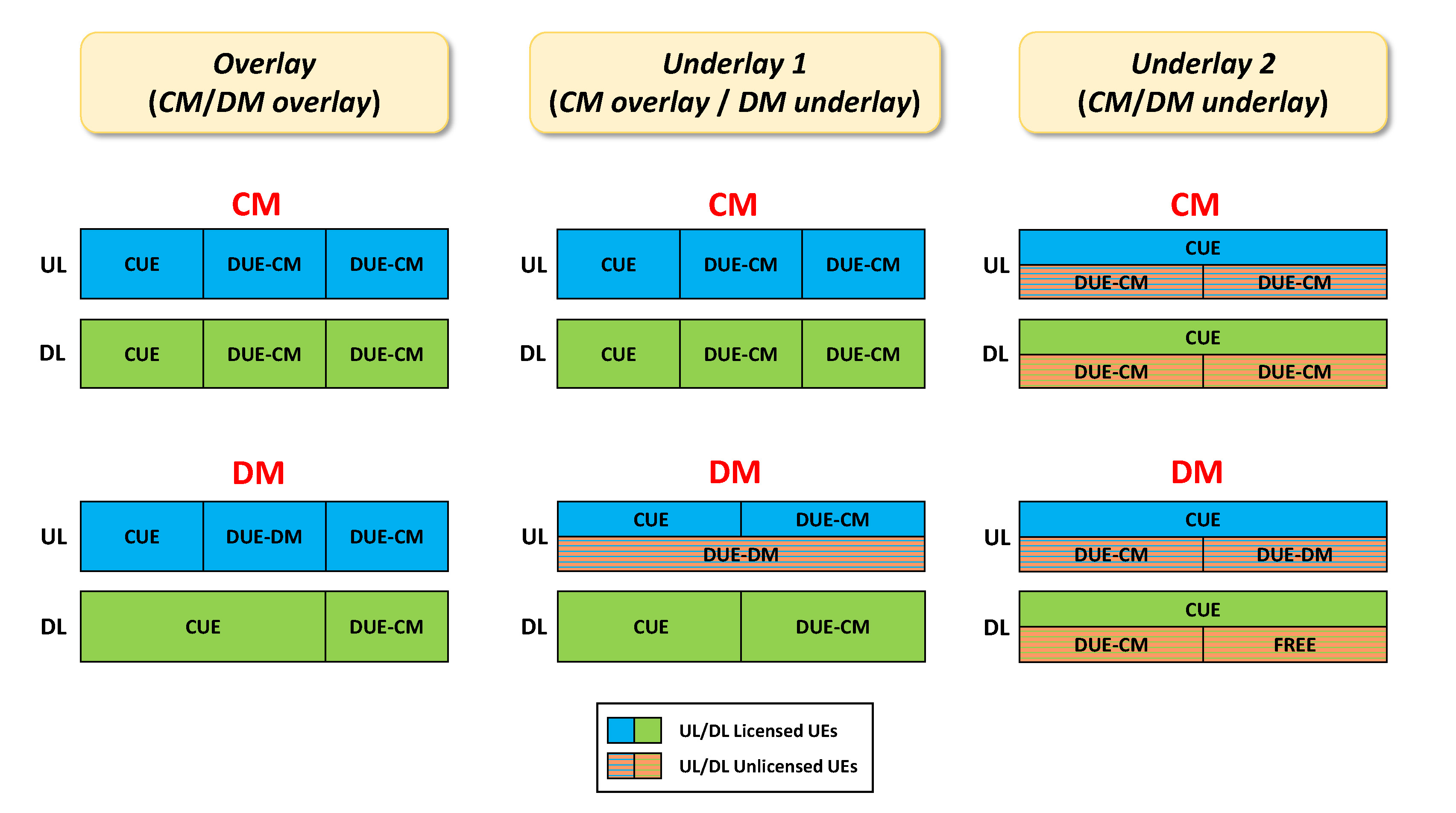}}
\caption{In-band spectrum sharing approaches.}
\label{fig2}
\end{figure}

\indent The above D2D communication modes utilize the licensed cellular spectrum either in an overlay or an underlay manner, while CUEs access the licensed spectrum in an orthogonal way. In particular, three D2D spectrum sharing approaches are considered: the {\it{Overlay}} approach, where potential DUEs communicate both for the CM and the DM utilizing the cellular resources orthogonally (CM/DM overlay or pure overlay); the {\it{Underlay 1}} approach, where the DUEs use orthogonal resources for the CM and non-orthogonal resources for the DM (CM overlay / DM underlay or mixed overlay/underlay); and the {\it{Underlay 2}} approach, where the DUEs utilize non-orthogonal resources both for the CM and the DM (CM/DM underlay or pure underlay). It is noted that the communication in DM utilizes the UL cellular resources, while the corresponding DL resources are offloaded in most of the cases for the legacy CUEs (except for some cross-cell D2D cases). Fig.~\ref{fig2} depicts the UL and DL resource utilization of the elaborate in-band spectrum sharing approaches.\newline
\subsection{Densification}
\indent Network infrastructure densification increases wireless network throughput and it is considered as a promissing solution for the booming traffic demand. Network densification can be defined as the deployment of more base stations and access points per unit area or volume. Specifically, the degree of densification can be characterized by the {\it{density}} or {\it{densification ratio}}. This ratio is defined as the ratio of access point density (number of eNBs per unit area) to user density (number of UEs per unit area).
According to the metric of density ratio, networks can be differentiated into sparse infrastructure deployments, with less access points than users and (ultra-) dense deployments, otherwise. However, combining network densification with D2D communications may not result in additive performance benefits, due to the fact that in more dense networks access points are placed closer to the users, but D2D communications are based on users proximity compared to their distance to access point. Consequently, the role of network densification in the D2D performance should be examined thoroughly when designing next generation cellular networks.

\section{D2D Radio Resource Management Mechanism}
\subsection{RRM Challenges \& Optimization Problem}
RRM determines the optimal use of wireless resources according to the current network status and the quality-of-service requirements of
the users. D2D-enabled system design faces various resource management issues, including mode selection, scheduling, channel quality estimation, power control, and beamforming. The biggest D2D RRM challenge is to efficiently incorporate D2D communications in a cellular network in a way the total network utility to be optimized (i.e. by ameliorating potential D2D pairs performance, without however degrading legacy CUEs performance). Most of the RRM research efforts have been focused on underlay single-cell spectrum sharing approached \cite{Xu}, while very few deal with overlay D2D or/and multiple cells \cite{Penda, Poulakis}. This article presents a joint overlay and underlay study in a multi-cell D2D-enabled environment. In particular, we focus on the design of D2D mode selection and UE scheduling procedures that constitute two of the most critical issues for the incorporation of D2D communications in legacy cellular systems. These decisions making procedures are highly intertwined, since scheduling is affected by the available UL and DL resources and the interference, which depend on mode selection of DUEs and vice versa. Consequently, these procedures should be handled jointly.\newline
\indent Specifically, a common entity, the resource manager, shares the resources over CUEs and DUEs at each time slot and at each cell and also decides the transmission mode (DM or CM) of each potential D2D pair. The proper UE should be carefully scheduled at each slot and at each cell, since the utilization of UL and DL resources depends also on the mode of D2D communication and the potential DL offloading. An example of the UL and DL resource allocation of two-phase communication (FDD or TDD) for CUEs, DUEs in CM and DUEs in DM is presented in Table~\ref{table1}.
\begin{table}[!h]
\centering
\caption{UL \& DL resource allocation.}
\label{table1}
\resizebox{0.9\width}{0.9\height}{
\begin{tabular}{@{}llll@{}}
\cmidrule[\heavyrulewidth](l){2-4}
& \multicolumn{3}{c}{\textbf{Scheduled UE}} \\ \midrule
\textbf{Phase} & \textbf{CUE} & \textbf{Pot. D2D pair} \textbf{in CM} & \textbf{Pot. D2D pair} \textbf{in DM}  \\ \midrule
\textbf{UL} & $CU{E_{{i_l}}} \rightarrow eNB_l$ & $DU{E_{{j_{1,l}}}}\rightarrow eNB_l$ & $DU{E_{{j_{1,l}}}}\rightarrow DU{E_{{j_{2,m}}}}$ \\
\textbf{DL} & $eNB_l \rightarrow CU{E_{{i_l}}}$ & $eNB_m \rightarrow DU{E_{{j_{2,m}}}}$ & FREE (offloaded)  \\ \bottomrule
\end{tabular}
}
\end{table}
\newline
\indent In order to design the D2D RRM mechanism, we formulate the joint mode selection and scheduling procedures as an optimization problem. A predetermined objective (utility function) must be optimized under constraints dictating the feasibility of the solution. The considered optimization variables are binary vectors $\pmb{x}\in\{0,1\}^{K^{pD2D}}$ and $\pmb{y}\in\{0,1\}^{K}$, determining the mode selection status of each DUE (DM or CM) and the scheduling status of each UE (both for CUE and DUE). Moreover, since the D2D pairs utilize UL resources, we are interested in optimizing the UL performance of the total system, taking into account at the same time the necessary allocations in the DL. Thus, the utility functions for CUEs (UL use), potential D2D pairs in DM (UL use) and potential D2D pairs in CM (UL and DL use) reflect the maximum instantaneous reliable rate of the corresponding links and are given by
\begin{footnotesize}
$$\boxed{
\begin{array}{l}
U_{{i_l}}^{leg} \;\;\;= \text{UL Capacity} \;(CU{E_{{i_l}}} \rightarrow eNB_l) \nonumber \vspace{1mm} \\
U_{{j_{l - m}}}^{DM} = \text{Pair Capacity} \; (DU{E_{{j_{1,l}}}}\rightarrow DU{E_{{j_{2,m}}}}) \nonumber \\
U_{{j_{l - m}}}^{CM} =  \min \left\{ \text{UL Capacity} \;(DU{E_{{j_{1,l}}}}\rightarrow eNB_l), \right. \nonumber \\
\;\;\;\;\;\;\;\;\;\;\;\;\;\;\;\;\;\;\;\;\;\;\;\;\; \left. \text{DL Capacity}\; (eNB_m \rightarrow DU{E_{{j_{2,m}}}}) \right\} \nonumber
\end{array}
}$$
\end{footnotesize}
Note that the utility function of potential D2D pairs in CM is defined as the maximum instantaneous reliable rate of the worst between UL and DL. \newline
\indent In the following, we present the joint problem for the different spectrum sharing schemes.
\subsubsection{Overlay Spectrum Sharing (CM/DM Overlay)}
\indent In {\it{Overlay}} spectrum sharing, the DUEs communicate both for the CM and the DM mode using the dedicated cellular resources. In this case, the joint optimization problem of overlay D2D mode selection and UE scheduling that aims at the maximization of total UL network performance in terms of utility function, in a multi-cell environment with intra-cell and cross-cell D2D communications enabled, can be formulated as follows
\begin{footnotesize}
$$\boxed{
\begin{array}{ll}
\mathop {\max }\limits_{\pmb{x},\pmb{y} } & U_{total}(\pmb{x},\pmb{y}) = f(U_{{i_l}}^{leg}, U_{{j_{l - m}}}^{DM}, U_{{j_{l - m}}}^{CM})\nonumber \\
{\text{ s.t.}} \,\,\,\,\,\,\, & (\text{C}1):\, \text{maximum number of UEs in UL} \nonumber\\
&(\text{C}2): \, \text{maximum number of DUEs in DL} \;\;\;\;\;\;\;\text{(P1)} \\ \label{P1}
&(\text{C}3):\,  \text{mode \& scheduling variables relation} \nonumber\\
&(\text{C}4):\,  \text{D2D distance restriction} \nonumber
\end{array}
}$$
\end{footnotesize}
where $U_{total}$ is the total utility function to be optimized which is a weighted sum of $U_{{i_l}}^{leg}, U_{{j_{l - m}}}^{DM}$ and $U_{{j_{l - m}}}^{CM}$. This function should be defined carefully in order to cover the cross-cell D2D cases. The (C1) corresponds to constraint of maximum simultaneous scheduled UEs at UL (one CUE or one DUE regardless the communication mode), while (C2) ensures that no more than one DUE can occupy a specific DL resource in cell $l$ (i.e. the total DUE transmitters in CM located in all the rest cells that have pairs in cell $l$  are no more than one). Furthermore, (C3) refers to the relation between mode and scheduling variables, while (C4) restricts the distance of a DUE pair to a maximum value in order to be able to communicate directly. Moreover, note that the employed scheduling scheme can be included in the optimization problem as an extra weight factor of each UE (CUE or DUE) (e.g. for round robin this factor reflects the reciprocal of each UE's scheduled times, while for proportional fairness it reflects the reciprocal of each UE's average capacity).

\subsubsection{Underlay Spectrum Sharing 1 (CM Overlay / DM Underlay)}
According to the {\it{Underlay 1}} spectrum sharing, the DUEs use dedicated resources for the CM and shared resources for the DM. Consequently, the DUEs in DM cause intra-cell interference to legacy CUEs and vice versa. Thus, the corresponding joint optimization problem of D2D mode selection and UE scheduling, can be formulated similarly to (P1) with the only modification the substitution of constraint (C1) with a new one allowing one CUE or one DUE in CM, and one DUE in DM simultaneously scheduled at UL of the same cell.

\subsubsection{Underlay Spectrum Sharing 2 (CM/DM Underlay)}
According to the {\it{Underlay 2}} spectrum sharing, the DUEs use shared resources both for the CM and the DM. Thus, the DUEs in all communication modes (CM or DM) cause intra-cell interference to legacy CUEs and vice versa. In this case, a common or a separate resource manager could be employed to decide about the mode selection and scheduling processes. Similar to Underlay 1, the corresponding joint optimization problem of D2D mode selection and UE scheduling, can be formulated by substituting constraint (C1) with a new one that grants spectrum access to one CUE and one DUE (regardless the communication mode) at the same cell, simultaneously.\newline
\indent It it noted that the above optimization problems are integer (or more precisely binary) non-linear and non-convex programs, which are difficult to be solved and to provide global optimal guarantees even for small-scale setups. To address this issue, we first reformulate the initial problem to a linear form, by eliminating the nonlinearities\footnote{For more details regarding the linearization-reformulation see \cite{Poulakis},\cite{Hou}.}. The transformed formulation is a binary (0-1) integer linear programming problem that is NP-complete and may be solved using integer linear programming employing popular and well-known solvers (e.g. GUROBI,GLPK, CPLEX) and derive a global optimal solution.

\subsection{Mode Selection \& Scheduling Policy}
\begin{figure}[!h]
\centering{
\hspace*{-3mm}
\frame{\includegraphics[width=3.0 in, scale=1]{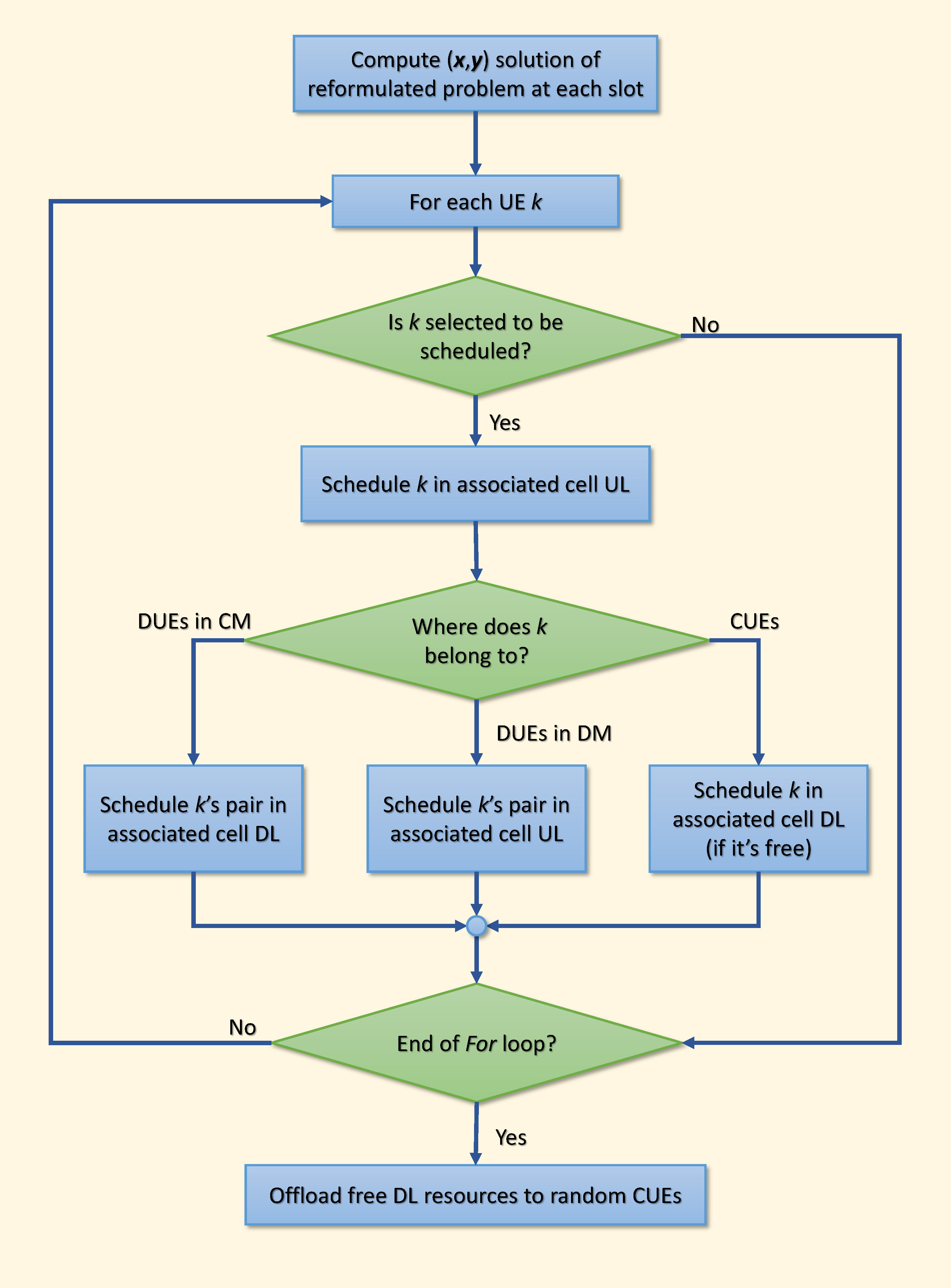}}}
\caption{Flowchart of mode selection and scheduling procedures.}
\label{fig3}
\end{figure}

Therefore, according to the joint mode selection and scheduling policy of the proposed RRM mechanism, a centralized resource manager should solve the joint optimization problem and the solution will indicate the UE that will be scheduled at each round and at each cell and moreover the communication mode (directly or through the eNB) if the UE belongs to potential D2D pairs. Fig.~\ref{fig3} depicts the followed joint mode selection and scheduling procedure in a flowchart form determining each transmission snapshot of the entire network.
%

\section{D2D Performance Analysis}
\subsection{System-level Simulation Results}
To evaluate the presented RRM algorithm and investigate the trends of D2D communications performance in relation to network and user densification, as well as the spectrum sharing schemes, we develop a system-level D2D-enabled cellular simulator. Realistic assumptions are considered, summarized in Table~\ref{table2}. Five cell-types with different radius sizes are assumed, resulting in the same total network coverage area and reflecting different densification ratios. Note that cell-type 1 refers to a single-cell scenario with the particularity of absence of cross-cell D2D communications. Regarding the interference, the multi-cell frequency reuse-1 is assumed for the legacy cellular network, while depending on the D2D spectrum sharing approach, inter-cell and intra-cell interference occurs. For consistency purposes, a simulated interference cumulative distribution function has been derived to generate random interference to boundary cell edges with no adjacent cells (including the cell-type 1 case). Furthermore, the eNB and UE antenna gains as well as the transmit powers are determined depending on the cell-type (i.e. eNBs’ density) in order to maintain a specific SNR value at cell-edges, considering a path loss exponent equal to 3. The optimization problem is modeled in MATLAB using the CVX software and solved using the GUROBI optimizer. Finally, the distance limit for D2D communications  is assumed to be equal to cell radius. \newline
\begin{table}[!t]
\caption{System-level simulator parameters}
\centering 
\resizebox{0.85\width}{0.85\height}{
\begin{tabular}{@{}cccccc@{}}
\toprule
\textbf{Parameter}                                                          & \multicolumn{5}{c}{\textbf{Value}}                                                                            \\ \midrule
Spectrum sharing                                                            & \multicolumn{5}{c}{In-band; Overlay / Underlay 1 / Underlay 2}
\\
D2D communication scenario                                                  & \multicolumn{5}{c}{In-coverage; same or different cell D2D}
\\
System bandwidth (MHz)                                                      & \multicolumn{5}{c}{5}                                                                                \\
Carrier frequency (GHz)                                                     & \multicolumn{5}{c}{2.6}                                                                              \\
\begin{tabular}[c]{@{}c@{}}Total network coverage area ($\rm{km}^2$)\end{tabular} & \multicolumn{5}{c}{0.234}                                                                            \\
Noise figure (dB)                                                           & \multicolumn{5}{c}{7}                                                                                \\
\begin{tabular}[c]{@{}c@{}}Noise spectral density (dBm/Hz)\end{tabular}  & \multicolumn{5}{c}{−174}                                                                             \\
Traffic model                                                               & \multicolumn{5}{c}{Full buffer}                                                                      \\
Scheduler type                                                              & \multicolumn{5}{c}{Round Robin}                                                    \\ \midrule
Cell type                                                                   & Type 1              & Type 2             & Type 3           & Type 4            & Type 5            \\
Cell radius (m)                                                             & 300                 & 212                & 150                & 123                & 100                \\
\begin{tabular}[c]{@{}c@{}}Network layout (number of cells)\end{tabular} & 1                   & 2                  & 4                 & 6                 & 9                 \\
eNB density (number/km$^2$)                                                 & 4.3               & 8.5              & 17.1            & 25.6              & 38.5              \\
eNB max power (dBm)                                                         & 23.0               & 21.0              & 20.0            & 17.4              & 14.7              \\
eNB antenna gain (dBi)                                                      & 3.0                 & 1.8                & 0.0              & 0.0                 & 0.0                 \\
UE max power (dBm)                                                          & 23.0               & 21.0              & 20.0            & 17.4              & 14.7             \\
UE antenna gain (dBi)                                                       & 3.0                 & 1.8                & 0.0              & 0.0                 & 0.0                   \\
\bottomrule
\end{tabular}
}
\label{table2}
\end{table}
\begin{figure}[!t]
\centering{
\hspace*{-12mm}
\includegraphics[width=4.3in, scale=1]{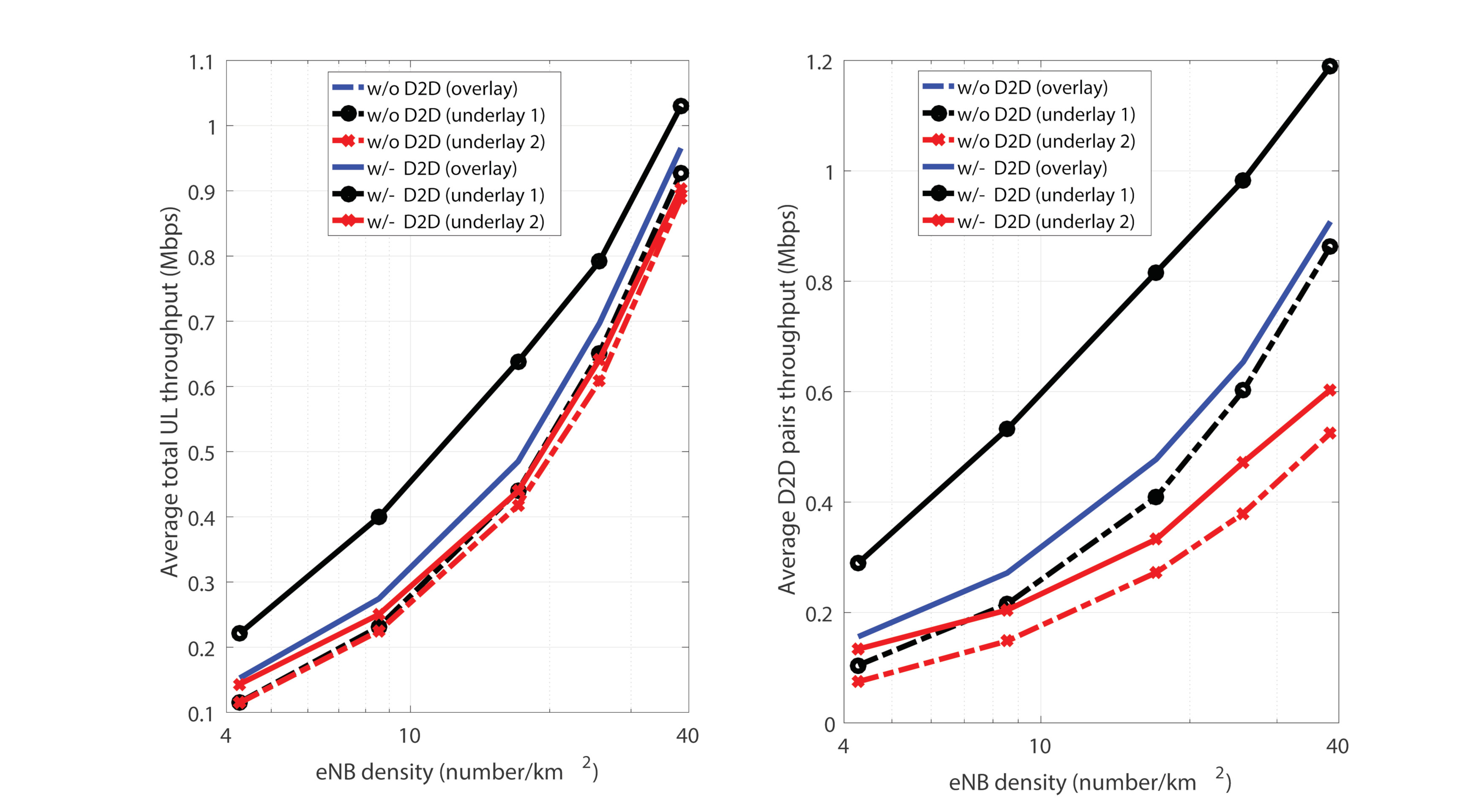}}
\caption{Average throughput versus eNB density for 108 UEs in total.}
\label{fig4}
\end{figure}
\indent  D2D communications can improve system's performance through the proximity gain, the hop gain and the reuse gain. In the considered scenario, where D2D communication utilizes UL resources, proximity gain can ameliorate UL performance, hop gain can increase DL performance, while reuse gain can improve both coming with the cost of interference. In Fig.~\ref{fig4}, the average total UL throughput of the network (i.e. UL throughput of CUEs and DUEs) is illustrated for the case of D2D communications disabled (i.e. all DUEs in CM) and the case of D2D communications enabled (i.e. the chosen DUEs by mode selector in DM) considering the different spectrum sharing approaches for increasing eNB density. As we can observe, the total UL throughput increases for higher densities, demonstrating the benefits of densification. Moreover, D2D communications increase the network's UL performance due to the proximity gain (enjoyed only by DUEs) with the improvement to be larger in the Underlay 1 approach due to the extra reuse gain (enjoyed both by DUEs and CUEs), while in Underlay 2 the reuse gain is offset by the caused interference. It is noted that the curves of Overlay and Underlay 1 w/- D2D are identical. Moreover, in order to focus on the improvement in DUEs throughput, Fig.~\ref{fig4} illustrates the average throughput of D2D pairs (derived from the corresponding utility functions) for the same scenarios. From this subfigure, we can observe similar results regarding the behaviour of the throughput, however the benefit is greater since the DUEs' throughput is isolated where the reuse (in underlay cases) and proximity (in all cases) gains are combined. \newline
\indent Furthermore, in order to evaluate the performance gain of D2D communications, we define the metric of total D2D communication gain ${\cal G}_{tot}$. This metric is defined as the weighted sum of direct communication gain ${\cal G}_{dir}$ in UL and the gain obtained by DL offloading ${\cal G}_{off}$. In particular, the direct gain (derived by proximity and reuse gain when it exists) corresponds to the percentage of throughput gain of potential D2D pairs, comparing the throughput achieved with D2D communications enabled and its absence. Moreover, the offloading gain [derived by hop gain (or reuse gain in Underlay 2 case)] that is obtained through D2D communications corresponds to the percentage of DL throughput gain of legacy CUEs comparing the case of coexistence with the potential D2D UEs with D2D communication enabled and the case where D2D communication is disabled (all the DUEs in CM). Furthermore, the total gain can be given by
$$\boxed{
\begin{array}{l}
{\cal G}_{tot}\, = {a_1} \cdot {\cal G}_{dir} + {a_2} \cdot {\cal G}_{off}
\end{array}
}$$
\begin{table}[!t]
\centering
\caption{D2D communications gain contribution.}
\label{tableG}
\resizebox{0.85\width}{0.85\height}{
\begin{tabular}{@{}lcccccc@{}}
\midrule
\begin{tabular}[l]{@{}l@{}}\bf{Spectrum}\\\bf{Sharing}\end{tabular} & \multicolumn{2}{c}{{Overlay}} & \multicolumn{2}{c}{{Underlay 1}} & \multicolumn{2}{c}{{Underlay 2}} \\
\midrule
\bf{Gain} & ${\cal G}_{dir} $ & ${\cal G}_{off} $  & $ {\cal G}_{dir} $ & ${\cal G}_{off} $ & ${\cal G}_{dir} $  & $ {\cal G}_{off} $  \\
\midrule
\bf{Proximity} & \ding{51}                                 &                                            & \ding{51}                                      &                                                  & \ding{51}                                  &                                              \\
\bf{Hop}       &                                             & \ding{51}                                &                                                  & \ding{51}                                      &                                              &                                              \\
\bf{Reuse}    &                                             &                                            & \ding{51}                                      &                                                  & \ding{51}                                  & \ding{51}                                  \\ \bottomrule
\end{tabular}
}
\end{table}
where $a_1$ and $a_2$ are the weight factors that corresponds to the considered contribution of each gain and without loss of generality are assumed to be equal to 0.5.\newline
\indent Table~\ref{tableG} summarizes the contribution of proximity, hop and reuse gain to the direct and DL offloading gain for the different spectrum sharing approaches. It is noted that in Underlay 1 and Underlay 2 approaches, an extra UL offloading gain is derived by reuse gain with the cost of interference, however this is out of the scope of this article.\newline
\indent In Fig.~\ref{fig5}, we present the average direct gain and offloading gain that the D2D communications achieve for different network cell-type topologies (see Table~\ref{table2}) assuming 108 UEs in total (CUEs and DUEs) and considering the same coverage area. The different cell-types represent different number of eNBs per same coverage area with increasing densification. As we can observe, the direct gain decreases in all cases as the network densification increases, because in more dense networks the distance of UEs to the associated eNB becomes smaller and thus the direct communication is more rarely preferred. This indicates the {\it{tradeoff}} between the performance improvement derived by {\it{densification}} and the {\it{D2D}} performance enhancement. We can also see that in general Underlay 1 spectrum sharing has better results due to extra reuse gain in DM, which gain is almost compensated by the interference in Underlay 2 scheme, while Overlay spectrum sharing leads to worst performance. Furthermore, similar results are observed for the offloading gain regarding network densification, while the hop gain of Overlay and Underlay 1 approaches is replaced by reuse gain in Underlay 2 approach. We have to note that, the hop gain always exists as only one hop is needed in DM instead of two hops in CM, nevertheless, both the proximity and reuse gains largely depend on the UEs locations.\newline
\begin{figure}[!t]
\centering{
\hspace*{-3mm}
\includegraphics[width=3.7in, scale=1]{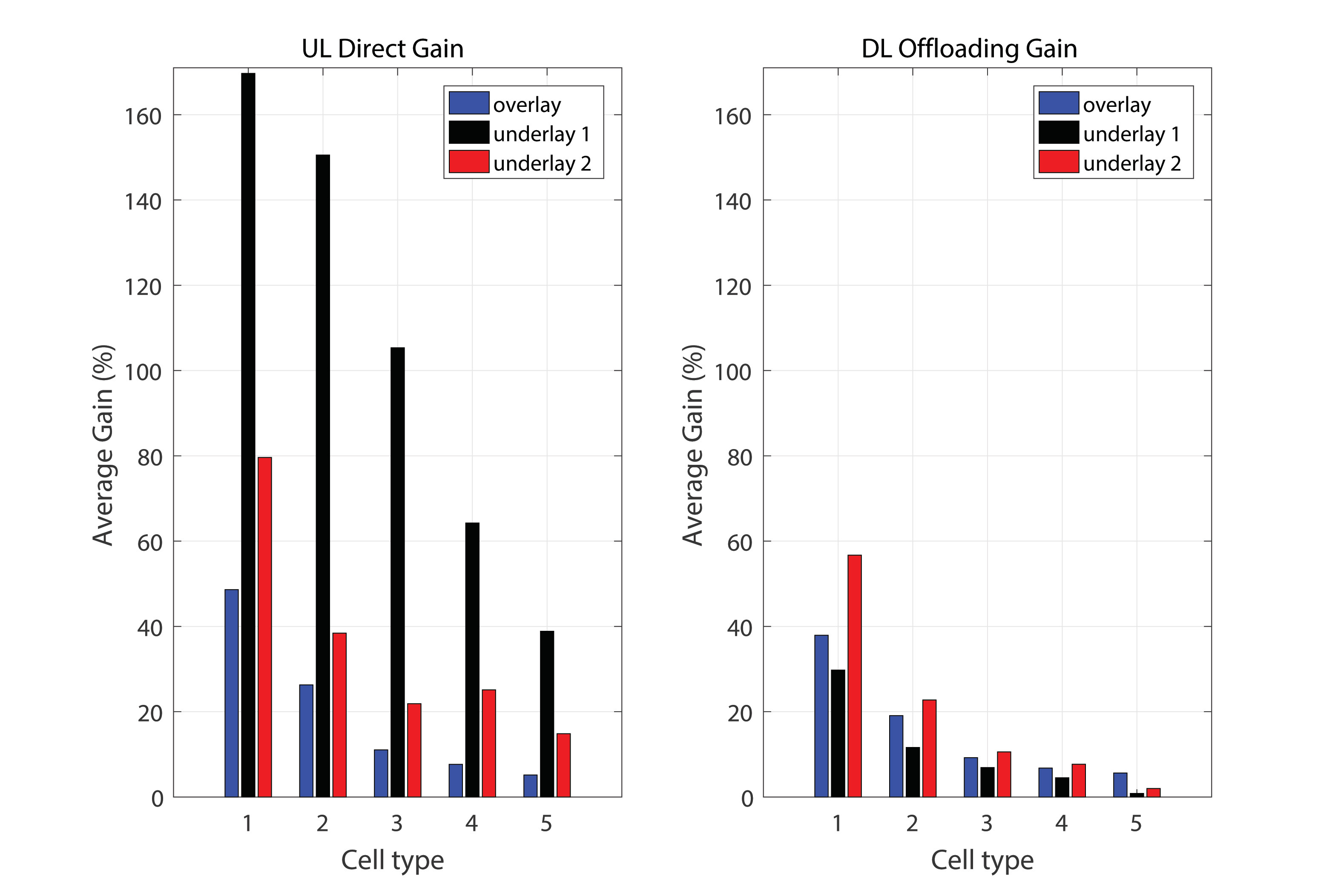}}
\caption{Average direct and offloading gain versus the different cell-type topologies (i.e. different eNB's densities).}
\label{fig5}
\end{figure}
\indent Finally, Fig.~\ref{fig6} illustrates the average total gain of D2D communications versus the different network cell-type topologies (i.e. different eNBs densities) and the UEs density, assuming 36 legacy CUEs and variant number of potential D2D pairs. The case of Overlay spectrum sharing is assumed. Fig.~\ref{fig6} indicates that the D2D benefit becomes greater as the number of UEs increases, since more opportunities to employ D2D communication appear, resulting in slightly larger direct gain, as well as more offloaded opportunities for CUEs are generated. Furthermore, the total D2D gain decreases as the number of eNBs (deployed in the given geographic area) increases for the same number of UEs. It can also be observed that the contribution of direct gain gain to the total gain decreases for more dense network and for higher number of UEs. In conclusion, a tradeoff between D2D gain and densification benefit exists and should be carefully considered when a system with D2D capabilities is designed.
\begin{figure}[!t]
\centering{
\hspace*{-3mm}
\includegraphics[width=3.6in, scale=1]{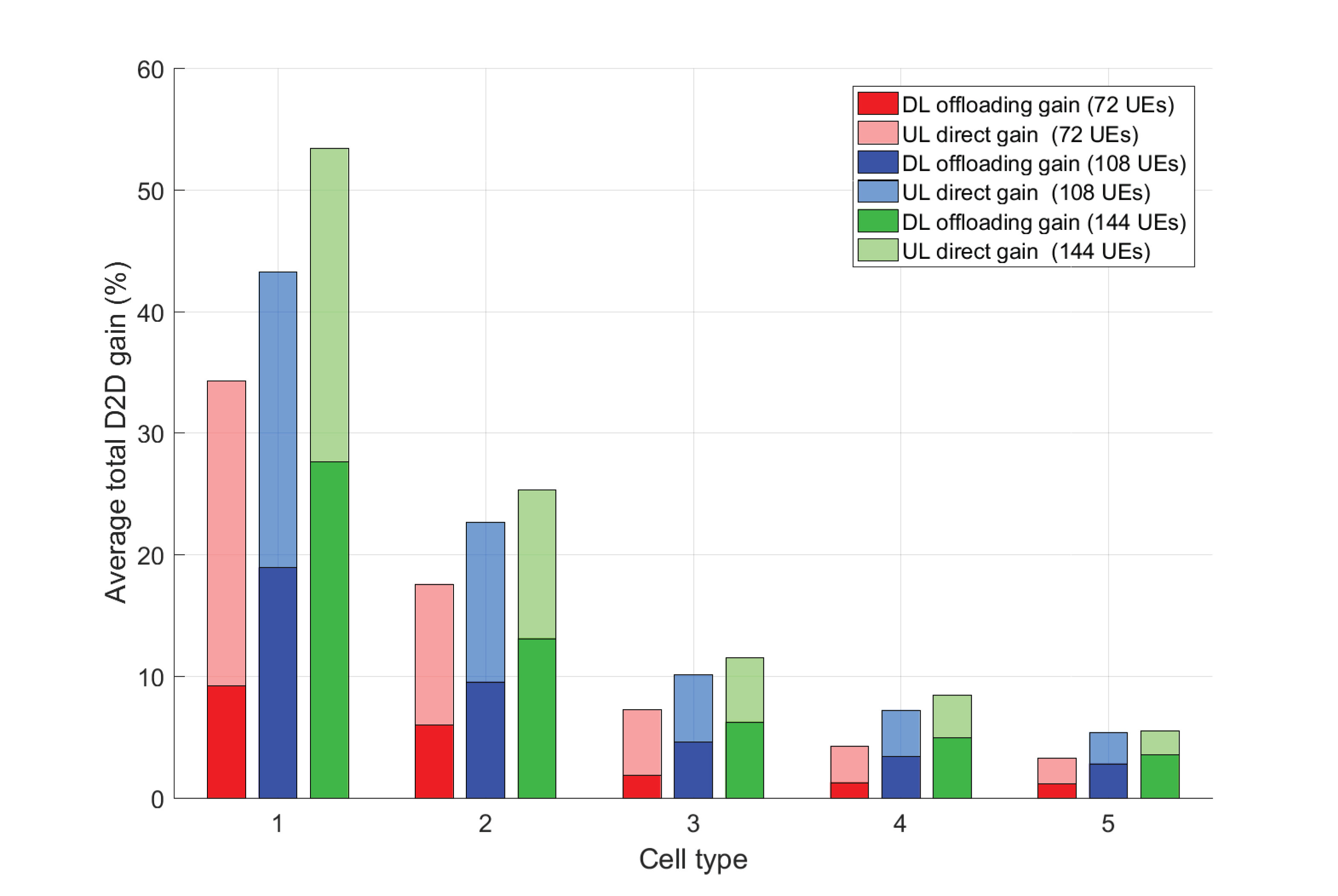}}
\caption{Average total gain of D2D communications versus the number of total UEs and different cell-type topologies for the Overlay approach.}
\label{fig6}
\end{figure}

\subsection{SDR-based Experimental Evaluation}
\begin{figure}[!t]
\centering{
\hspace*{-3mm}
\includegraphics[width=3.2in, scale=1]{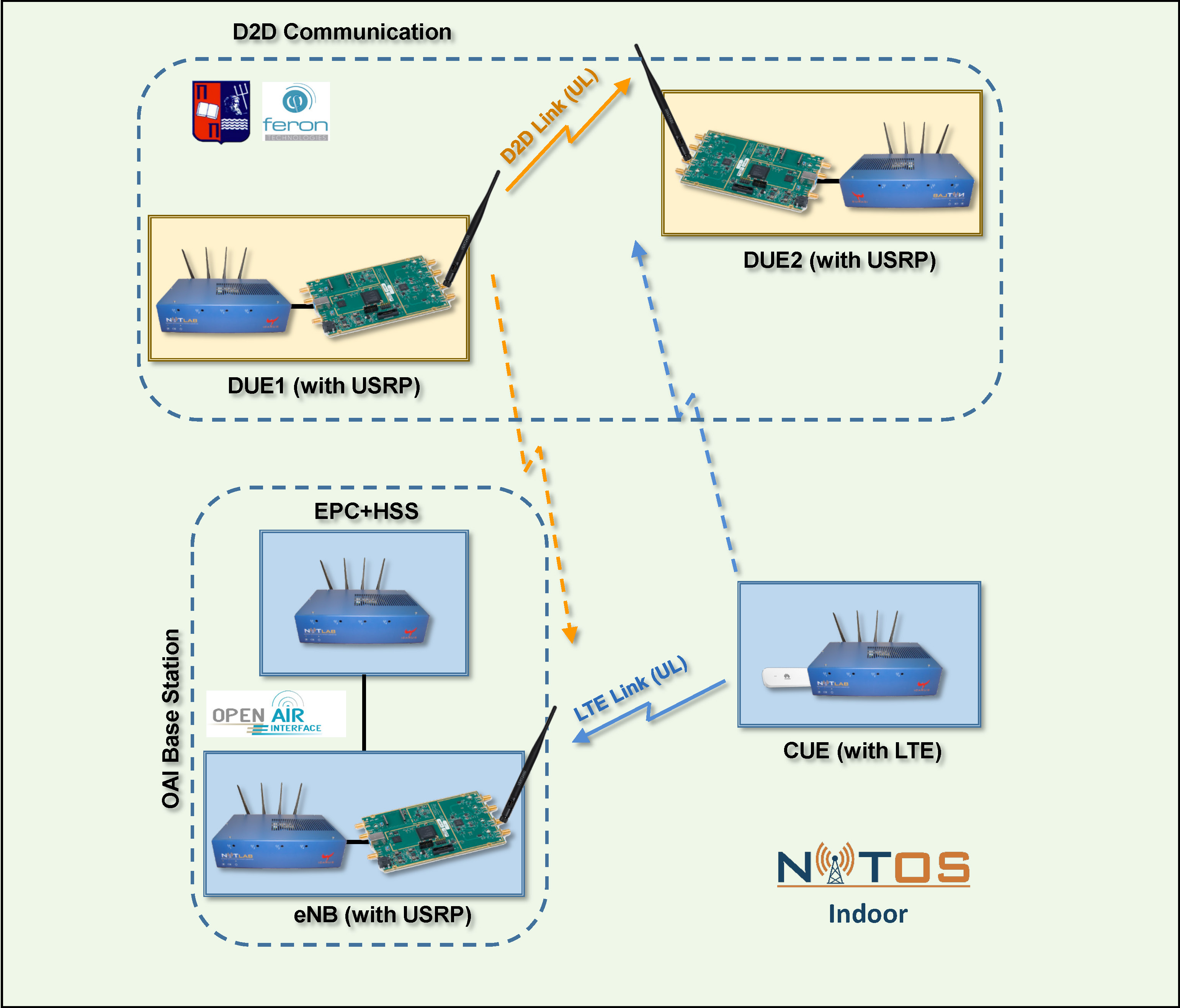}}
\caption{Joint LTE-D2D experimental setup.}
\label{fig7}
\end{figure}

\begin{figure}[!t]
\centering{
\hspace*{-3mm}
\includegraphics[width=3.0in, scale=1]{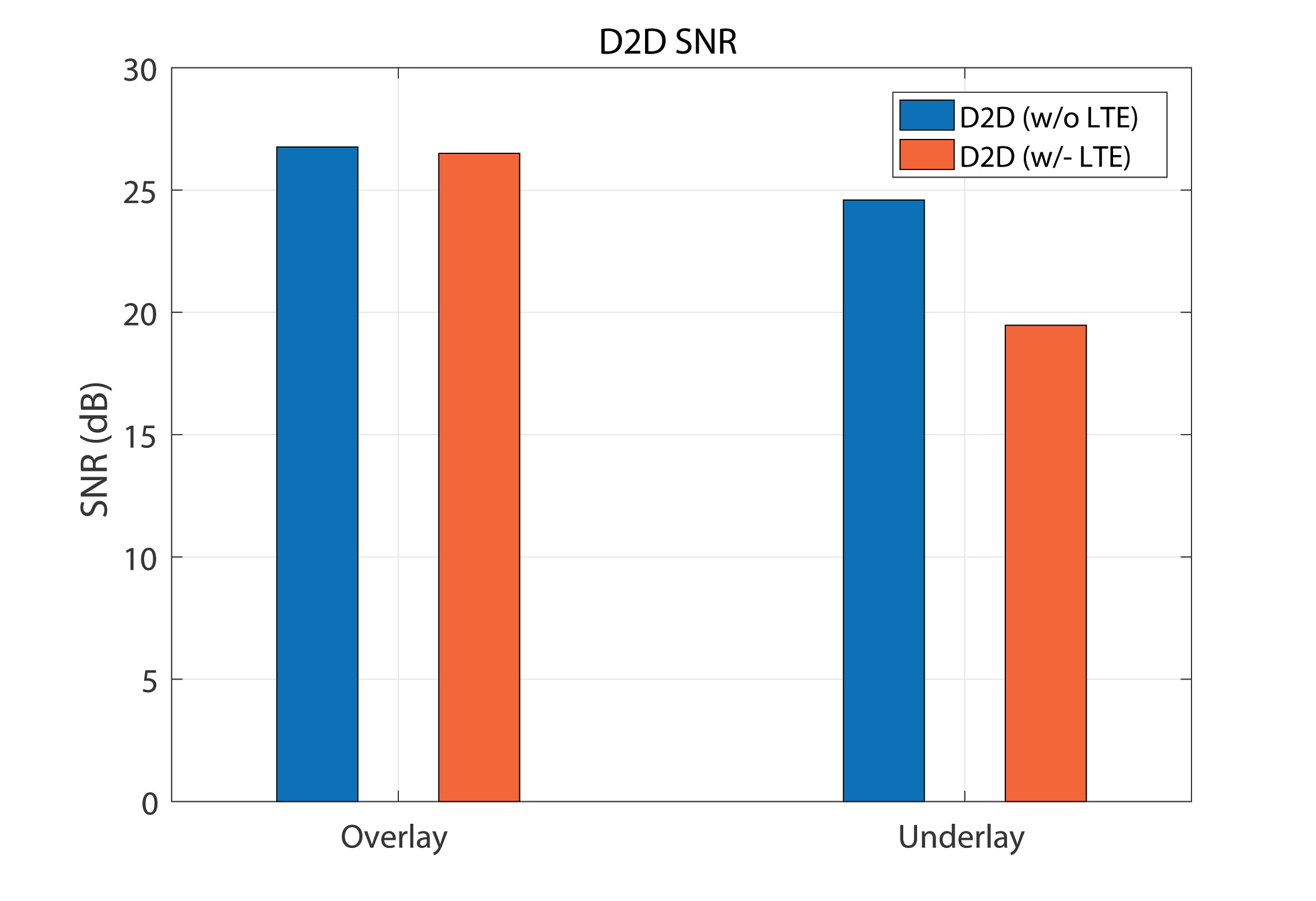}}
\caption{D2D experimental results for spectrum sharing schemes comparison.}
\label{fig8}
\end{figure}

\begin{figure}[!t]
\centering
\subfigure[LTE SNR]{
   \includegraphics[width=1.6in, scale=1] {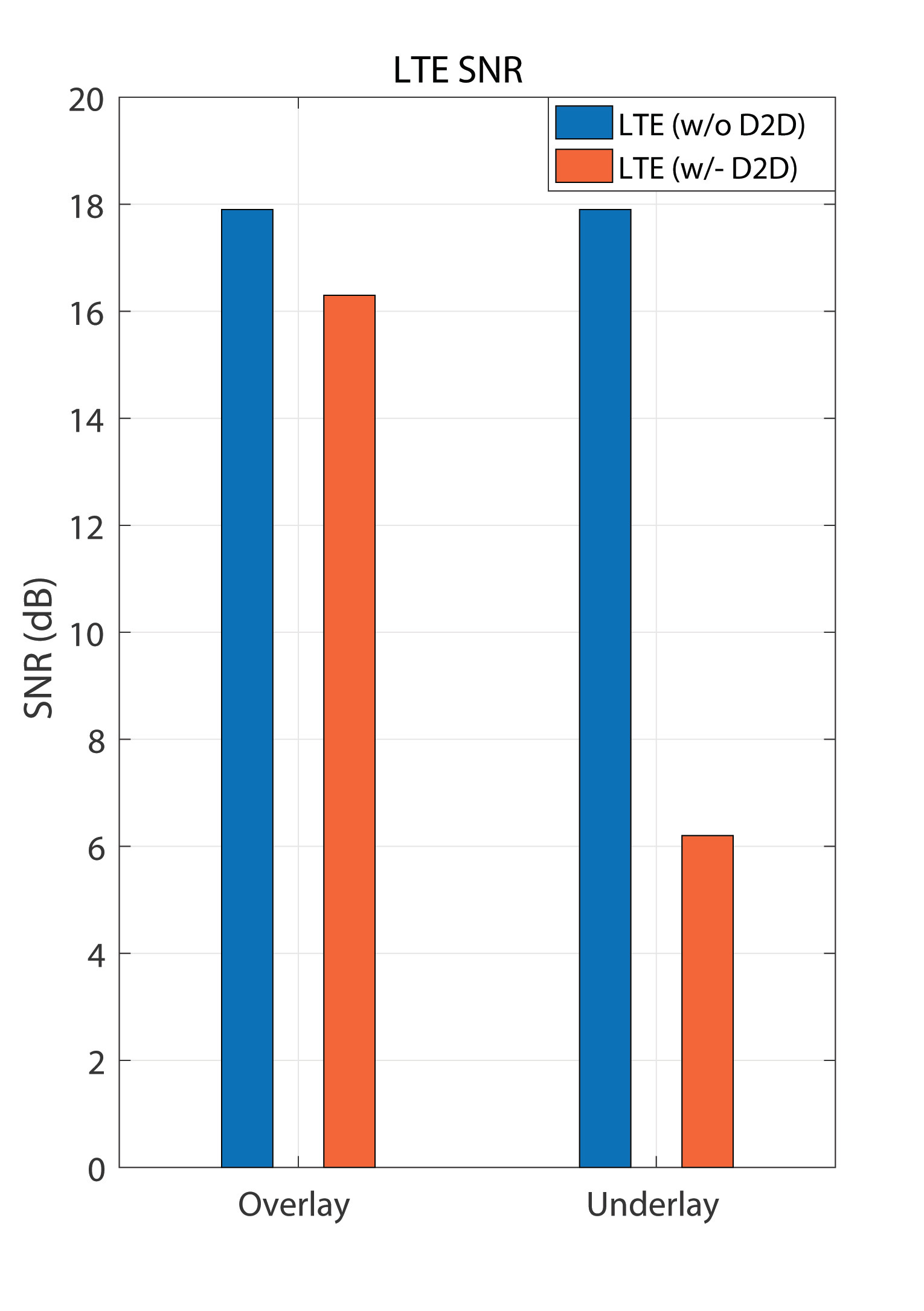}
   \label{fig9a}
 }
 \subfigure[LTE Throughput]{
   \includegraphics[width=1.6in, scale=1] {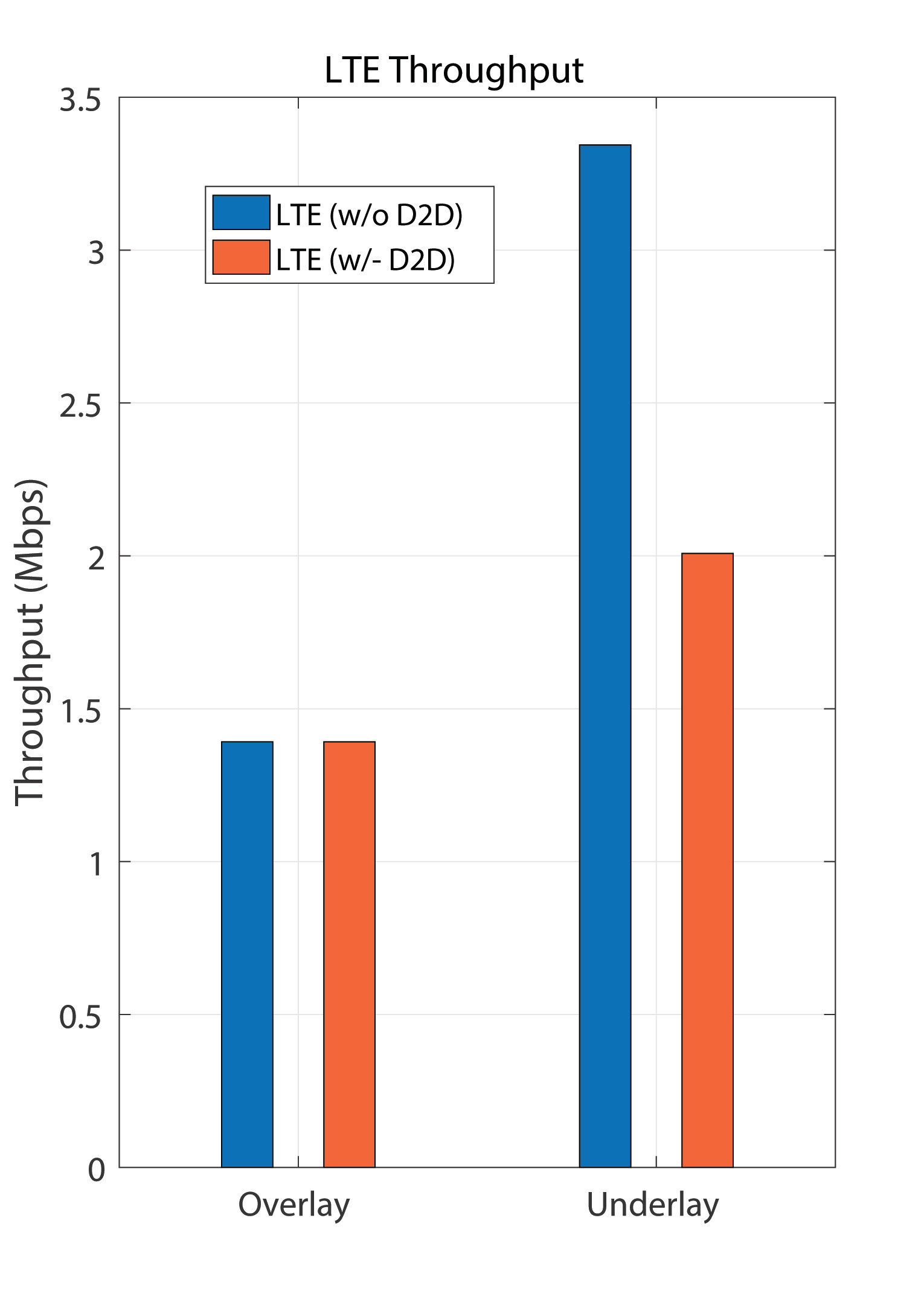}
   \label{fig9b}
 }
\caption{LTE experimental results for spectrum sharing schemes comparison.}
\label{fig9}
\end{figure}

This subsection is devoted to the experimental evaluation of in-band D2D communications in the testbeds of FLEX project \cite{FLEX}, which offers a valuable, flexible and credible solution for open and cost efficient LTE experimentation. In particular, we employ the NITOS indoor testbed \cite{NITOS} and the OpenAirInterface (OAI) platform \cite{OAI} in order to perform joint LTE-D2D tests and investigate the impact of D2D communication to a legacy LTE operation and vice-versa, for the different spectrum sharing schemes. A baseline coexistence setup is assumed, where an LTE pair and a D2D pair in DM communicate accessing the same UL spectrum in an overlay or an underlay manner. Specifically, the LTE pair consists of a $CUE$ node equipped with commercial LTE dongle and two SDR\footnote{Note that Ettus B210 USRP boards were employed.}-equipped nodes hosting the OAI core network code (EPC+HSS) and a modified version of OAI eNB code, respectively, for the implementation of LTE base station. Regarding the D2D pair, two SDR-equipped nodes implement the D2D transmitter ($DUE_1$) and receiver ($DUE_2$). Since the D2D pair is considered to transmit in DM, no mode selection procedure is employed and consequently the underlay case reflects both the Underlay 1 and Underlay 2 approaches that were described in the previous sections. EARFCN Band 3 (UL: 1.715GHz/ DL: 1.810GHz) is utilized, while the 5 MHz channel bandwidth case is considered. In the overlay case, D2D and LTE pairs share the available bandwidth in a fully orthogonal manner, utilizing a part of the total available bandwidth (i.e. 8 PRBs in total for each pair), while no interference arises among each other. In the underlay case, DUEs use the same resources for the DM transmission as CUE in a non-orthogonal manner using all the available bandwidth (i.e. 20 PRBs in total for each pair), however (possibly) harmful interference is generated among each other. \newline
\indent Fig.~\ref{fig7} illustrates the experimental setup (topology). It is worthwhile to mention that the nodes' locations were selected aiming at the ‘harmonious’ coexistence of LTE and D2D pairs in order to be able to perform overlay and underlay tests. This setup reflects near-cell-edge D2D communications, in which the D2D benefit is expected to be more significant.\newline
\indent Moreover, to assess the performance of the spectrum sharing scenarios, the following ‘co-existence’ tests are performed:
\begin{enumerate}
\item \textit{Interference-free LTE}: Legacy LTE UL transmission without D2D communications (D2D w/o LTE)
\item \textit{Joint LTE-D2D}: Simultaneous LTE-D2D UL transmission (D2D w/- LTE and LTE w/- D2D)
\item \textit{Interference-free D2D}: D2D UL transmission without Legacy LTE transmission (LTE w/o D2D)
\end{enumerate}
\indent With regard to the evaluation Key Performance Indicators (KPIs), we measured the following metrics in the UL, concerning both the legacy LTE UE and D2D UE: \textit{D2D SNR}, \textit{LTE SNR} and \textit{LTE throughput}. Note that the SNR values are extracted in an empirical manner by measuring the Error Vector Magnitude (EVM) of the received signals and using the following formula: $SN{R_{dB}} \approx 10 \cdot {\log _{10}}\left( {\frac{1}{{EV{M^2}}}} \right)$. \newline
\indent Fig.~\ref{fig8} and Fig.~\ref{fig9} present the comparative results for the spectrum sharing schemes considering the joint LTE-D2D system as well as the reference individual systems separately (interference-free). The highest D2D transmit power level (i.e. 3.35mW) corresponding to the highest value of interference caused to the LTE pair (just before OAI system breaks down), is assumed for all the experiments. More specifically, Fig.~\ref{fig8} depicts the D2D SNR measurements. As it can be observed, the reduction of SNR is negligible in overlay scenario compared with the SNR reduction in the underlay case, due to the orthogonal resource utilization. Note that the degradation would be larger for lower D2D transmit power levels.\newline
\indent Furthermore, Fig.~\ref{fig9} illustrates the corresponding LTE experimental results for the different spectrum sharing schemes. In particular, Fig.~\ref{fig9a} presents the SNR values, while Fig.~\ref{fig9b} shows the UL throughput measurements. As we can observe, the SNR degradation due to joint LTE-D2D operations is larger than that observed in the D2D pair, since the D2D interference is at its maximum point, for the underlay scenario. Moreover, for the overlay scenario, where there is no spectrum overlap (nor guard bands) and thus no interference is expected, a slight reduction is observed due to the adjacent-channel interference. Finally, in Fig.~\ref{fig9b}, we can observe a significant throughput reduction in the joint LTE-D2D operation compared with interference-free LTE setup for the underlay case. In contrast, for the overlay scenario, the throughput remains the same for the interference-free LTE and the joint LTE-D2D cases. It is worthwhile to mention that the underlay spectrum sharing exhibits higher throughput values, even after its significant degradation due to the D2D interference. If the interference increases more than this point, the LTE connection is expected to be lost in the underlay case, whereas the overlay LTE-D2D systems will maintain a slower on the one hand, but relatively stable connection on the other hand.\newline
\indent Consequently, the best spectrum sharing policy depends on the individual experiment conditions, with the underlay to be more appropriate for low-interference regimes providing faster connections (for restricted interference values), whereas the overlay scheme seems to be more suitable for high-interference regimes offering a little slower but stable and robust connections, regardless of the experienced interference.

\section{Conclusion and Future Challenges}
D2D communication and network densification are considered to be among the main enablers for 5G wireless systems. However, the increase of densification may affect the D2D performance in D2D-enabled systems. This article presented a study on the effects of densification and spectrum sharing on the D2D performance. A D2D RRM mechanism, which jointly optimizes mode selection and scheduling procedures in a multi-cell system with overlay/underaly D2D communication capabilities, was proposed. System-level simulations were performed to evaluate the proposed mechanism and examine the benefits of D2D communications for different system parameters. The results have shown that the D2D gain (consisting of direct and offloading gain) is significantly affected by the spectrum sharing scheme, where the Overlay scheme offers the worst performance since only proximity and hop gains exist, while as the reuse gain is introduced this leads to performance enhancement with the additional cost of interference. Furthermore, it was concluded that in all cases, the increasing densification comes at the expense of lower D2D gain. Therefore, a tradeoff between D2D benefit and densification gain was observed, indicating that direct communications are more beneficial in less dense scenarios. Moreover, real-world SDR-based experiments comparing the overlay and underlay schemes, constituting a proof-of-concept for the feasibility of in-band D2D communications, revealed that the best policy depends on the individual experiment conditions, with the underlay to be more appropriate for low-interference regimes providing faster connections, whereas the overlay schemes seem to be more suitable for high-interference regimes offering stable and robust connections. \newline
\indent There are many open challenges on the road to efficient D2D-enabled systems design apart from mode selection, scheduling, spectrum sharing and network densification. Indicatively, discovery process, energy efficiency, mobility management and security mechanisms need to be carefully addressed as well. Moreover, although D2D communication is heavily studied, research carried out so far is still in the preliminary stage of studying D2D performance in simplified scenarios. Finally, note that D2D communication can be viewed as a possible enabler of Vehicle to everything (V2x) communication that has attracted great interest due to the potential of improving traffic safety and enabling new intelligent transportation services.

\section*{Acknowledgment}
This work was funded by EU-FP7 ``FLEX'' Project (grant agreement 612050).

\section*{Author Information}
\noindent \textbf{Marios I. Poulakis} received the Diploma degree in electrical and computer engineering from the National Technical University of Athens (NTUA), Greece, and the M.Sc. degree in Management and Economics of Telecommunication Networks from the National \& Kapodistrian University of Athens (NKUA), Greece, in July 2006 and December 2008, respectively. In May 2014, he received the Dr.-Ing. degree from the NTUA. In 2007, he joined the Mobile Radio Communications Laboratory, NTUA as an associate researcher / project engineer participating in various industry and research-oriented projects in the telecommunications sector. Since 2016, M. Poulakis has been a post-doctoral researcher at the Department of Digital Systems, University of Piraeus, Greece. His research interests include wireless and satellite communications, as well as cognitive radio networks, with emphasis on optimization mechanisms for quality of service driven scheduling and resource management. He is a Member of the IEEE.\newline
\noindent \textbf{Antonis G. Gotsis} received his Diploma and Ph.D. degrees in electrical and computer engineering from the National Technical University of Athens (NTUA), Greece, in 2002 and 2010, respectively. From September 2002 to December 2010, he was with the Mobile Radio-Communications Laboratory, NTUA, as a research assistant/project engineer in various communications engineering projects. In 2012, he joined the Department of Digital Systems, University of Piraeus, Greece, where he worked as a principal researcher in a national research and development project. Since 2015, he is with Feron Techologies, working as R\&D Engineer and leading the company's research programmes. His research interests include radio resources management for advanced wireless systems, interference coordination techniques, application of optimization tools in wireless communications and networks, and prototyping and demonstrating air-interface system concepts in software-defined radio platforms. He is a Member of the IEEE.\newline
\noindent \textbf{Angeliki Alexiou} received her Diploma in electrical and computer engineering from the National Technical University of Athens, Greece, in 1994 and her Ph.D. degree in electrical engineering from Imperial College of Science, Technology, and Medicine, University of London, in 2000. Since May 2009, she has been a faculty member with the Department of Digital Systems, University of Piraeus, Greece, where she conducts research and teaches undergraduate and postgraduate courses in the area of broadband communications and advanced wireless technologies. She is currently an associate professor with the Department of Digital Systems, University of Piraeus, Greece. Prior to this appointment, she was with Bell Laboratories, Wireless Research, Lucent Technologies (now Alcatel-Lucent), Swindon, United Kingdom, first as a member of technical staff (January 1999-February 2006) and later as a technical manager (March 2006-April 2009). Her current research interests include radio interface and MIMO technologies, cooperation and coordination techniques and efficient radio resource management for ultradense wireless networks, and machine-to-machine communications. She is a Member of the IEEE.\\

\end{document}